\documentclass{svjour3}                     
\smartqed  
\usepackage{graphicx}
\begin{document}

\title{Latest results for the antikaon-nucleon optical potential
\thanks{Presented at the 21st European Conference on Few-Body Problems in Physics, Salamanca, Spain, 30 August - 3 September 2010.}
}


\author{V.K. Magas \and J. Yamagata-Sekihara \and S. Hirenzaki \and E. Oset \and A. Ramos
}


\institute{V.K. Magas \at
              Departament d'Estructura i Constituents de la Materia,\\ 
              Universitat de Barcelona, Diagonal 647, E-08028 Barcelona, Spain \\
              Tel.: +34-93-4039178\\
              Fax: +34-93-4021198\\
              \email{VLADIMIR@ECM.UB.ES}           
           \and
           J. Yamagata-Sekihara \at
             Yukawa Institute for Theoretical Physics, Kyoto University, Kyoto 606-8502, Japan {\LARGE $\&$}\\
       Departamento de F\'{\i}sica Te\'orica and IFIC, Centro Mixto Universidad de Valencia-CSIC,\\ Institutos de Investigaci\'on de Paterna, Apartado 22085, 46071 Valencia, Spain
           \and
           S. Hirenzaki \at
             Department of Physics, Nara Women's University, Nara 630-8506, Japan
           \and
           E. Oset \at
             Departamento de F\'{\i}sica Te\'orica and IFIC, Centro Mixto Universidad de Valencia-CSIC,\\ Institutos de Investigaci\'on de Paterna, Apartado 22085, 46071 Valencia, Spain
           \and
           A. Ramos \at
             Departament d'Estructura i Constituents de la Materia,\\ 
             Universitat de Barcelona, Diagonal 647, E-08028 Barcelona, Spain
}

\date{Received: date / Accepted: date}

\maketitle

\begin{abstract}
The key question of this letter is whether the $K^-$-nucleus optical potential is deep, as it is prefered by the phenomenological fits to kaonic atoms data,  or shallow, as it comes out from unitary chiral model calculations. The current experimental situation is reviewed. 
\end{abstract}

\bigskip
The issue of the antikaon interaction in the nucleus has attracted much attention in the last years. Although from the study of kaon atoms it is known for a long time that the $K^-$-nucleus potential is attractive, the discussion centers on how attractive the potential is and whether it can accommodate deeply bound kaonic states. 

Deep $K^-$-nucleus optical potentials are preferred in the phenomenological fits to kaon atoms data - the resulting potential can be up to 600 MeV deep in the center of the nucleus \cite{3}.  According to \cite{3} if the potential is deep enough to bind $K^-$ inside the nucleus so strongly that the main decay channels ($K^- N \rightarrow \Lambda \pi,\ \Sigma \pi$) will be blocked, this will open a new very rich field of physics - kaonic nuclei. These very compact and very dense objects, with central density which can be 10 or even more times larger than the normal nuclear density \cite{3}, could then be produced in the laboratories. These deeply bound $K^-$ states in the nuclei will be long living, and, thus, can be observed and directly studied experimentally, contrary to dense and hot, but highly dynamical systems created in relativistic heavy ion collisions. A sufficiently large attraction could also make possible the existence of kaon condensates in nuclei and in neutron stars.

On the other hand, the existence of such states would be a call for theoreticians to look for some modifications of chiral unitary models, because all potentials based on underlying chiral dynamics of the kaon-nucleon interaction are shallow (of the order of 40-60 MeV attraction at nuclear matter density) and with large imaginary part (also about 40-60 MeV at nuclear matter density) \cite{4,5}. Such potentials do produce $K^-$ bound states in nuclei, but these cannot be observed experimentally, because their width is much larger than the energy separation between the levels.

\begin{figure*}
  \includegraphics[width=0.63\textwidth]{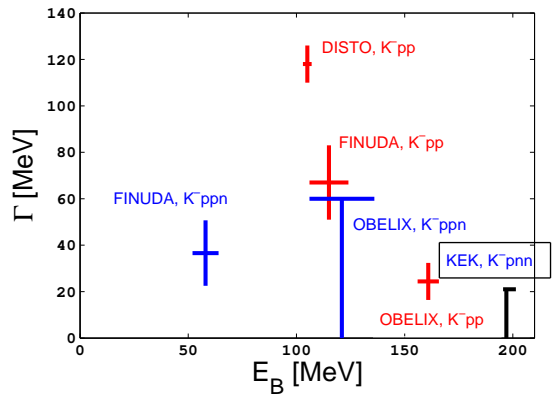}
\caption{Experimental claims of the observed deeply bound $K^-$ states with nucleons: $K^-$pp clusters from FINUDA \cite{13}, OBELIX \cite{17}, and DISTO \cite{18}; and $K^-$ppn clusters from KEK \cite{11}, FINUDA \cite{14}, and OBELIX \cite{17}. }
\label{fig:2}       
\vspace{-0.5cm}
\end{figure*}

From the experimental side the search for bound $K^-$ states with nucleons is a most direct and clear way to answer whether the $K^-$-nucleon potential is deep or shallow, because only a deep potential may generate states sufficiently narrow to be observed experimentally. Since 2004 several claims of observed deeply bound $K^-$ states have been made, but the situation is still very unclear  (see Fig. \ref{fig:2}).

The first claim of $K^-$pnn bound state came from the $K^-_{stop}+ ^{4}He \rightarrow p+X$ proton missing mass experiment at KEK \cite{11}, and it was withdrawn after the new more precise experiment \cite{12}. 
Later came two claims from the FINUDA experiment of the observation of $K^-$pp bound state in $\Lambda$p invariant mass spectra in stopped $K^-$ reaction with different light nuclei \cite{13}; and of the observation of $K^-$ppn bound state in $\Lambda$d invariant mass spectra in stopped $K^-$ reaction with  $ ^{6}Li$ \cite{14}. However, in \cite{1,2,15,16} we have shown that all the experimental signals used to support these claims of the observed deeply bound $K^-$ clusters with nucleons can be explained (within the experimental accuracy) without bound states by means of $K^-$ absorption from a low lying atomic orbit by two  or three  nucleons correspondingly, leaving the rest as spectators; and correctly taking into account the 
nuclear medium effects, namely the Fermi motion/recoil for the light nuclei like $ ^{4}He$ \cite{16}, $ ^{6}Li$ \cite{2}; 
and	the final state interaction of the particles, generated in $K^-$ absorption, with daughter nucleus \cite{1}. 

There are also claims of $K^-pp$ and $K^-ppn$ bound states from $\bar{p}$ annihilation in $ ^4He$ at rest measured by OBELIX@CERN \cite{17}, however their statistical significance is very low. The most recent is the claim of $K^-pp$ bound state, seen in $pp\rightarrow K^+ X$  reaction, from DISTO experiment \cite{18}. These experimental claims are under investigation now.  Before calling in new physics one has to make sure that these data cannot be explained with conventional mechanisms. 

One thing can be noticed immediately by looking at Fig. \ref{fig:2} - the binding energies and width of the same cluster from the claims of different experiments do not agree with each other. Thus, there is no way that all these experiments are right at the same time. Thus, there is a 
clear call for a reanalysis of the experimental data.

We believe that carefully reanalysing the existent data we can learn a lot about $K^-$ two and three nucleon absorption mechanisms, about final state interactions, etc. This knowledge will be a key element of the future analysis of new experiments with higher statistics, like E15 at J-PARC, AMADEUS at DA$\Phi$NE. 

We followed this strategy in \cite{1,2,22}; and now we are using the obtained knowledge in analysis of experiment data from \cite{19}. This experiment studied the $(K^-,p)$ reaction on $^{12}C$ with fast kaons with 1 GeV/c momentum. 
Our analysis \cite{Junko} shows that although this experiment is not suitable to extract information about the strength of the $K^-$ -nucleon optical potential, it can provide very clean information about two nucleon absorption of $K^-$ . 

\vspace{-0.5cm}
\begin{acknowledgements}
This work is partly supported by
the contracts FIS2008-01661 from MICINN
(Spain), by CSIC and JSPS under the Spain-Japan research Cooperative program,
 and by the Ge\-ne\-ra\-li\-tat de Catalunya contract 2009SGR-1289. We
acknowledge the support of the European Community-Research Infrastructure
Integrating Activity ``Study of Strongly Interacting Matter'' (HadronPhysics2,
Grant Agreement n. 227431) under the Seventh Framework Programme of EU.
\end{acknowledgements}



\end{document}